\documentclass[twocolumn,floats,floatfix,superscriptaddress,aps,pra]{revtex4}

\usepackage{eurosym}
\usepackage{booktabs} 

\usepackage{amsfonts}
\usepackage{amssymb}
\usepackage{amsmath}
\usepackage{graphicx}
\usepackage{lipsum} 
\usepackage{bm}
\usepackage{color}
\usepackage{xcolor}
\usepackage{epsfig}
\usepackage{ifthen}
\usepackage[section]{placeins}
\usepackage{array} 
\usepackage{subfigure}
\usepackage{physics}
\usepackage{floatrow}
\usepackage[hidelinks, colorlinks=true,linkcolor=blue,citecolor=blue,filecolor=blue,urlcolor=blue]{hyperref}

\begin{document}

\title{Robust, fast and high-fidelity
composite single-qubit gates for superconducting transmon qubits}

\author{Hristo G. Tonchev}
\affiliation{Center for Quantum Technologies, Department of Physics, Sofia University, James Bourchier 5 blvd., 1164
Sofia, Bulgaria}
\author{Boyan T. Torosov}
\affiliation{Institute of Solid State Physics, Bulgarian Academy of Sciences, 72 Tsarigradsko chauss\'{e}e, 1784 Sofia, Bulgaria}
\author{Nikolay V. Vitanov}
\affiliation{Center for Quantum Technologies, Department of Physics, Sofia University, James Bourchier 5 blvd., 1164
Sofia, Bulgaria}

\begin{abstract}
We introduce a novel quantum control method for superconducting transmon qubits that substantially outperforms conventional techniques in precision and robustness against coherent errors. Our approach leverages composite pulses (CP) to effectively mitigate system-specific errors, such as qubit frequency and anharmonicity variations. By utilizing CP, we demonstrate both complete and partial population transfers between qubit states, as well as the implementation of two essential single-qubit quantum gates. Simulations reveal substantial reductions in common error rates and gate durations. The effectiveness of our method is validated through four independent verification techniques, underscoring its potential for advancing quantum computing with superconducting qubits.

\end{abstract}

\author{}
\maketitle

\section{Introduction}
\label{introduction}
Quantum computing (QC) promises to revolutionize computer science by introducing an astonishing paradigm shift in how we think about information. It has been shown that, for certain problems, quantum algorithms significantly outperform the best known classical analogues \cite{nielsen2010quantum}. Until recently, quantum computing was, to a large extent, concentrated in academic institutions and was considered mainly an interesting scientific exercise with unclear potential for 'real-world' applications. This was mainly due to the difficulties in scaling up quantum computers to a large number of qubits. These difficulties themselves lead to the development of new types of algorithms (e.g., VQE \cite{Peruzzo_2014,McClean_2016}, QAOA \cite{farhi2014quantumapproximateoptimizationalgorithm}) that are aimed at
achieving useful results by utilizing quantum computers, which are currently available. Quantum computing entered the so-called NISQ (Noisy Intermediate-Scale Quantum) era \cite{Preskill2018quantumcomputingin}.

With the expansion of quantum computing from academia to industry came a (maybe temporary) shift in the focus on the physical systems, used to represent
the qubits. While in academic institutions the interest is spread among a variety of qubit systems, including
trapped ions, atoms, photons, quantum dots, etc., in commercially available quantum computers, the clear front runner is the superconducting qubit, mostly due
to fabrication facilitation. The most popular type of superconducting qubit is the so-called transmon, which is essentially an LC circuit, where the linear inductance
is replaced by a Josephson junction \cite{Koch_2007}. For its lowest states, such a system can be described as an anharmonic
oscillator, where the lowest two energies represent the qubit. The anharmonicity prevents from exciting the higher states, referred to as leakage errors, if the driving
fields are long enough, such that their Fourier spectrum does not cover the leakage transition.

The control over transmon qubits, used to produce quantum gates and algorithms, is  performed by
external microwave fields. Through this manipulation, several types of errors can arise \cite{Willsch_2017}, such as decoherence \cite{Burnett_2019}, population leakage outside the computational subspace \cite{Wood_2018,PhysRevA.94.032321}, and coherent (unitary) errors \cite{greene2021errormitigationstabilizermeasurement}.
Among these, the most crucial limitation
derives from leakage of population outside of the computational subspace, which limits the gate duration. The limitation stems from the Fourier bandwidth of a pulse: if the pulse is too short, then its Fourier bandwidth is too large and contains frequency components close to resonance with unwanted upper
transitions. The probability of such a leakage can be reduced by increasing the gate duration (and hence squeezing the Fourier bandwidth), by pulse shaping, active leakage cancellation, or most often, by using the so-called DRAG pulses \cite{DRAG, Dragog, theis2018counteracting,chiaro2025activeleakagecancellationsingle}, which can be seen as an extension of the popular “shortcuts to adiabaticity” method \cite{shortcut}. Furthermore, different extension of the DRAG method have been developed, such as pulse shaping in the frequency domain \cite{hyyppa2024reducing}, or using optimal control \cite{werninghaus2021leakage}.
In most cases, the idea is to eliminate the frequency components near resonance with the leakage transitions of higher energy levels. In two-qubit gates, leakage can be reduced by using a suitable synchronization of control parameters \cite{Barends_2019}, or by suitable pulse shaping, such as using Slepian pulses \cite{martinis2014fast}.

In this work, we introduce a different approach to eliminate the probability for leakage even with (previously prohibitively short) pulses with
frequency spectrum covering the transition. We use the technique of composite pulses (CP), which allows to enhance or reduce a certain probability generally at
will. The composite pulses are sequences of pulses with well-defined relative phases, which are used as control parameters to distort the excitation profile (and even the entire propagator) in a desired way. In the present context the leakage transitions are strongly suppressed by the destructive interference enabled by the composite sequence, even though every single pulse in the sequence produces non-negligible excitations to the upper states.

In particular, we demonstrate how two of the most common sources of errors in a superconducting qubit, leakage and control inaccuracies,
can be mitigated by a single unified approach. Decoherence is also addressed by shortening the gate duration, which is usually associated with a higher population leakage, due to the increased Fourier bandwidth. Furthermore, in line with Ref.~\cite{lagemann2023fragility}, which demonstrates the fragility of gate errors under pulse parameter variations, our work introduces a robust CP strategy that significantly mitigates these vulnerabilities while still maintaining good fidelity.
Below, we elaborate on our approach and present simulations demonstrating the method's performance.

\section{DESCRIPTION OF THE METHOD}
The transmon qubit is described by the Hamiltonian \cite{Koch_2007}
\begin{equation}
    \hat{H}_{\mathrm{tr}} = 4 E_C \,\hat{n}^2 \;-\; E_J \cos \hat{\phi},
    \label{eq:Hamiltonian tr}
\end{equation}
where \(E_C\) and \(E_J\) denote the capacitive and Josephson energies, respectively, and \(\hat{n}\) and \(\hat{\phi}\) are the reduced charge and phase operators satisfying the commutation relation \([\hat{\phi}, \hat{n}] = i\). Because of the cosine nonlinearity in the effective potential, the eigenenergies deviate from those of a simple harmonic oscillator, which becomes more pronounced at higher energy levels.

To numerically simulate the qubit, we first use the charge basis representation of the operators, where
\begin{equation}
  \begin{aligned}
    \hat{n} \;&=\; \sum_{n=-\infty}^{\infty} n \,\ket{n}\!\bra{n}, \\ 
    \cos \hat{\phi} \;&=\; \sum_{n=-\infty}^{\infty} \frac{1}{2} \bigl(\ket{n+1}\!\bra{n} \;+\; \ket{n-1}\!\bra{n}\bigr),
  \end{aligned}
  \label{eq: n_sum}
\end{equation}
with \(n\) representing the number of Cooper pairs on the island and \(\ket{n}\) the associated eigenstate of the charge operator. The expression for \(\cos\hat{\phi}\) follows directly from the canonical commutation relations \cite{Koch_2007}. For practical simulations, the summation over \(n\) must be truncated: we take \(n_{\text{min}} = -n_{\text{cut}}\) and \(n_{\text{max}} = +n_{\text{cut}}\), with \(n_{\text{cut}} = 30\) deemed sufficient to achieve the desired accuracy.

After constructing the Hamiltonian in the charge basis, we diagonalize it to obtain its representation in the energy eigenbasis. Next, we incorporate a drive term to enable single-qubit control, which is typically implemented by an external microwave voltage source coupled capacitively to the qubit. Hence, an additional term is introduced in the Hamiltonian, proportional to the charge operator
\begin{equation}
    \hat{H}_d = \Omega(t)\,\cos\bigl(\omega_d t + \phi\bigr)\,\hat{n},
    \label{eq:controled ham}
\end{equation}
where \(\Omega(t)\) is the (potentially) time-dependent Rabi frequency, \(\omega_d\) is the drive frequency, and \(\phi\) is its phase.
 
By moving to a rotating frame at the drive frequency and applying the rotating-wave approximation (RWA), the rapidly oscillating components in the drive term~\eqref{eq:controled ham} can be eliminated, leaving only terms with the slowly varying envelope~\(\Omega(t)\). A more detailed derivation of the model can be found in Appendix~\ref{app: A}.

Finally, we arrive at the complete (driven) Hamiltonian in the rotating frame
\begin{equation}
    \hat{H}(t)  = \sum_{j = 0}^{2n_{\text{cut}}} \mu_j \,\Pi_j \;+\;  
\sum_{j=1}^{2n_{\mathrm{cut}}}
\lambda_{j} \,\frac{\Omega_{R}(t)}{2}\,\ket{j}\!\bra{j-1} + \text{h.c.},
\label{eq:ham}
\end{equation}
where \(\mu_j = e_j - j\,\omega_d\) (with \(e_j\) denoting the transmon energies), \(\lambda_{j}\) are numerically derived parameters satisfying \(\lambda_{j}^{*} = \lambda_{j}\), and \(\Omega_{R}(t) = \Omega(t)\,\mathrm{e}^{-i\phi}\).


Having established a model for our controllable transmon, we now introduce our quantum control method. For a time-independent Hamiltonian, \(\Omega(t) = \Omega\), the system evolves according to the propagator
\[
    \mathbf{U} = \exp\bigl(-i\,\hat{H}(\Omega,\phi)\,T\bigr),
\]
where \(\hat{H}(\Omega,\phi)\) is the time-independent Hamiltonian from Eq.~\eqref{eq:ham}, and \(T\) is the total duration for which the propagator is applied. To address the problem of  off-resonant excitations to states outside of the computational basis, we employ composite pulses (CP). Rather than a single pulse described by \(\mathbf{U}\), we apply a sequence of \(N\) pulses, each with its own relative phase \(\phi_k\), Rabi frequency \(\Omega_k\), and gate duration \(t = T/N\). The total propagator is then given by
\begin{equation}
    \mathbf{U}^{(N)} \;=\; \mathbf{U}(\Omega_N,\phi_N,t)\,\cdots\,\mathbf{U}(\Omega_2,\phi_2,t)\,\mathbf{U}(\Omega_1,\phi_1,t).
    \label{eq: propag}
\end{equation}

We use the relative phases and Rabi frequencies as control parameters to minimize the leakage and shape the excitation profile in a desired robust fashion. We achieve this goal either by maximizing a certain transition probability or by maximizing the fidelity of a target gate. This is then extended over a certain range of deviations in the Rabi frequencies, while trying to keep the overall pulse area $\sum_{k=1}^N \Omega_k T/N$ as small as possible. For a more detailed discussion on composite pulses, we refer the reader to the vast literature on the topic \cite{Vandersypen_2005,freeman1997spin,soton352479}.



\section{Simulations}

To illustrate the performance of our method, we carry out two distinct types of simulations. The first focuses exclusively on the populations of the two lowest energy states after the interaction, thereby neglecting any phase information in the probability amplitudes. The second simulates the generation of quantum gates, where phases are essential—and hence the calculations become considerably more demanding. Therefore, we analyze these two cases separately.

The simulations are performed as follows. We calculate the propagator~\eqref{eq: propag}
for an \(N\)-pulse sequence, where we allow for a common systematic deviation \(\epsilon\)
in the Rabi frequencies. Consequently, each pulse is described by
\(\Omega_k (1 + \epsilon)\,\mathrm{e}^{i\phi_k}\). Our objective is then to determine
the parameters \(\Omega_k\) and \(\phi_k\) such that a desired excitation profile, across
a specified range of \(\epsilon\) values is achieved.

Before proceeding with the two types of composite sequences, we note that our optimization procedures assume a qubit frequency of \(\omega = 2\pi \times 7\)\,GHz and an anharmonicity of \(\delta = -2\pi \times 0.3\)\,GHz. The total evolution time is \(T = 20\)\,ns, divided into \(N\) independent control pulses. After computing \eqref{eq:ham}, we truncate the Hilbert space to the first six energy levels; our numerical simulations suggest that this number sufficiently captures transient leakage during evolution, and increasing it further does not alter the results. Additionally, we observe that reducing \(T\) below 20\,ns leads to composite-pulse sequences that violate the RWA criterion \cite{ZEUCH2020168327}.

The optimization of all numerical parameters is performed using the \texttt{scipy.optimize} package \cite{2020SciPy-NMeth}, specifically employing the standard L-BFGS-B optimizer \cite{zhu1997algorithm}. Moreover, whenever a time-dependent Hamiltonian is required (as in the case of DRAG simulations \cite{DRAG}), we use the Python library QuTiP \cite{JOHANSSON20131234}, which allows for the calculation of propagators for time-dependent \(\Omega(t)\). All of the code required to reproduce the results is available on GitHub \cite{megit}.

\subsection{Population Transfer}

\begin{table}

\footnotesize
\begin{tabular}{llc}
\hline
Target    & $(\Omega_1, \Omega_2, \cdots \Omega_N) $ & $  \qquad\mathcal{A}$ \\
& $(\phi_1,  \phi_2, \cdots \phi_N)$& \\
\hline
$P_1 = 1$      &  $(42.497,69.996,69.996,69.761,63.782,$ & $ \qquad 7.98$ \\ &
$69.996,58.263)$ &  \\ & $(-0.3875,0.0188,0.0191,0.1258,0.2469$ & \\ & $0.3139,0.2516)$  & \\ 
$P_1 = \frac{1}{2}$      &  $(27.127,38.559,42.584,38.388,32.701,39.706,$ & $ \qquad 3.88$ \\ &
$13.643,14.55)$ &  \\ & $(0.2202,0.066,-0.0306,0.0617,0.0889,0.1198,$ & \\ &
$0.1188,0.0152)$ &  \\  
 \hline
$X$      &  $(31.651,44.988,69.97,60.608,66.029,68.771,$ & $  \qquad 7.52
$ \\ &
$69.562,66.971)$ &  \\ & $(0.1779,0.0499,0.1239,0.2538,0.2886,0.1688,$ & \\ &
$0.1645,0.1234)$ &  \\  

$\sqrt{X}$      &  $(27.127,38.559,42.584,38.388,32.701,39.706,$ & $  \qquad 3.88
$ \\ &
$13.643,14.55)$ &  \\ & $(.2202,0.066,-0.0306,0.0617,0.0889,0.1198,$ & \\ &
$0.1188,0.0152)$ &  \\ 

\lasthline
\end{tabular}

\caption{Rabi frequencies $\Omega_{i}$
(in units $2 \pi \times $MHz) and relative phases $\phi_{i}$ (in units $\pi$) for the $N \in
\mathbf{N}$ optimization parameters $i\in{1,N}$ used for the precise
composite sequences, producing specific
transition probability (top part) or specific quantum gate
(bottom part). The total pulse area is denoted by $\mathcal{A}$.}
\label{table: params}
\label{Table1}
\end{table}

\begin{figure}[tb]
	\centering 
	\includegraphics[width=0.8\columnwidth, angle=0]{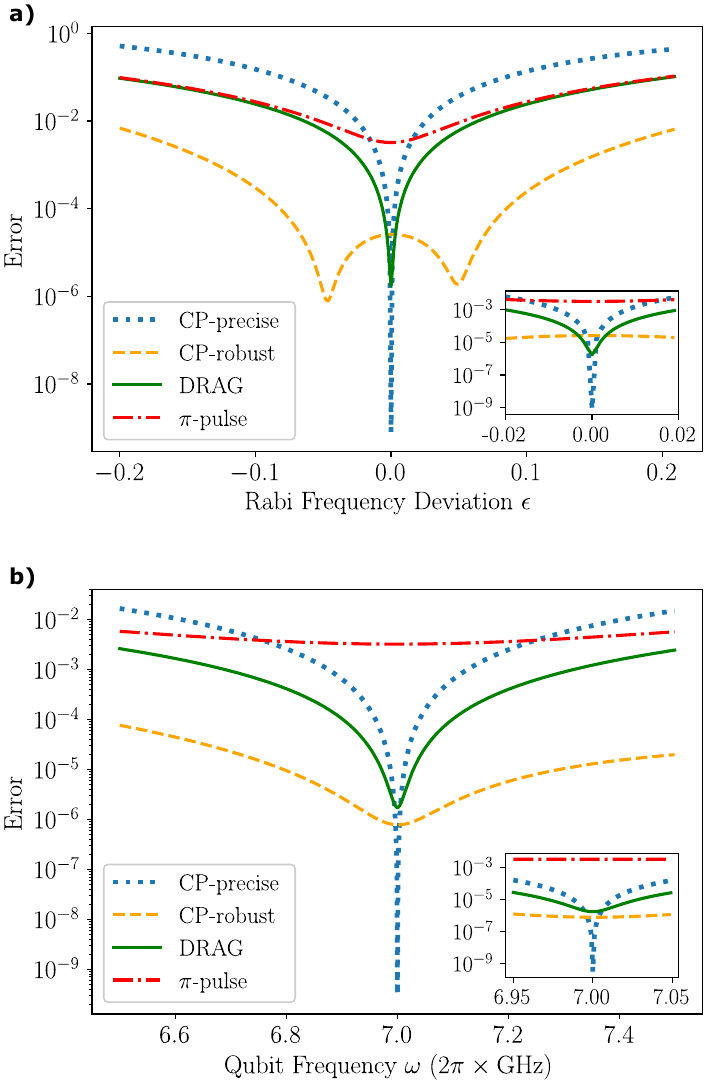}	
	\caption{(a) Population error in state $\ket{1}$ as a function of Rabi frequency error $\epsilon$. Results are shown for a single $\pi$-pulse, a precise composite sequence, a robust CP, and a DRAG-shaped pulse optimized to maximize transition probability. The inset zooms in on the region around $\epsilon = \pm 0.02$.
(b) Population error in state $\ket{1}$ as a function of qubit frequency $\omega$. All four pulse types from (a) are included. For the robust CP, we have assumed $\epsilon = -0.047$; for the others, $\epsilon = 0$. Pulse parameters, including composite phases and Rabi frequencies, are listed in Table~\ref{table: params}. Optimization settings are $N = 7$, with $\epsilon_{\text{max}} = 0$ for the precise solution and $\epsilon_{\text{max}} = 0.07$ for the robust one.
In both panels, the precise solution is highly sensitive to the system parameters used during optimization, indicating a need for accurate prior calibration. In contrast, the robust CP maintains better performance across a wider range of errors, trading accuracy for greater resilience to parameter variations.
}
	\label{fig: pop_trans}
\end{figure}

In the population transfer simulations, we assume the system is initially in the state \(\ket{0}\). Our objective is to achieve a specified transition probability in a robust manner. In particular, we consider two cases: complete population transfer, where \(P_1 = 1\), and half population transfer, where \(P_0 = P_1 = \tfrac{1}{2}\), with $P_j$ representing the population in state $j$. We determine optimal values for the parameters \(\{\Omega_k\}\) and \(\{\phi_k\}\) by minimizing the cost function
\begin{equation}
   f(\{\Omega_k\},\{\phi_k\}) 
   = \sum_{\epsilon = -\epsilon_{\max}}^{\epsilon_{\max}} 
         \sum_{i=0}^{1}
         \bigl|\bigl|U^{(N)}_{i1}(\Omega_k (1 + \epsilon),\phi_k)\bigr|^2 - P_i\bigr|,
   \label{eq: Cost Function}
\end{equation}
where the summation over \(\epsilon\) increases robustness to small deviations in the Rabi frequencies. We will explore how the choice of \(\epsilon_{\max}\) influences the trade-off between accuracy and robustness.

\begin{table}[ht]
\centering

\textbf{Robust Gates Errors} \\[2pt]
\begin{tabular}{|l|c|c|c|c|}
\hline
Gate & Rot. Error & Phase Error & Leakage & $1-F$ \\
\hline

X         & $8.62 \times 10^{-5}$ & $8.40 \times 10^{-8}$ & $1.5 \times 10^{-4}$ & $1.67 \times 10^{-4}$ \\
$\sqrt{X}$ & $3.73 \times 10^{-6}$ & $4.38 \times 10^{-10}$ & $2.68 \times 10^{-6}$ & $5.87 \times 10^{-6}$ \\
\hline
\end{tabular}

\vspace{1em}

\textbf{Precise Gates Errors} \\[2pt]
\begin{tabular}{|l|c|c|c|c|}
\hline
Gate & Rot. Error & Phase Error & Leakage & $1-F$ \\
\hline
X         & $1.39 \times 10^{-8}$ & $2.04 \times 10^{-10}$ & $2.75 \times 10^{-8}$ & $2.77 \times 10^{-8}$ \\
$\sqrt{X}$ & $2.44 \times 10^{-9}$ & $1.54 \times 10^{-10}$ & $4.23 \times 10^{-9}$ & $4.67 \times 10^{-9}$ \\
\hline
\end{tabular}

\caption{Comparison of error metrics for robust and precise implementations of $X$ and $\sqrt{X}$ CP gates. In all cases, rotation error and leakage constitute the dominant sources of infidelity, typically exhibiting comparable magnitudes. Phase error, by contrast, remains negligible across all gates.}
\label{table: split_gates}
\end{table}

One can also add an additional regularization term to Eq.~\eqref{eq: Cost Function} that pushes the optimization algorithm towards regions with a smaller pulse area. Such an approach though turns out to be computationally heavy for large systems, and instead picking out by hand solutions with smaller pulse areas is sufficient.

Our first simulation concerns the case of complete population transfer. Results for two types of solutions, that differ in robustness, are shown in Fig.~\ref{fig: pop_trans}. Both of them are compared to a standard $\pi$ and DRAG pulses. For the latter the first order $Y$-only correction is used (known as simple DRAG \cite{DRAG}), optimized to minimize Eq.~\eqref{eq: Cost Function}, and with an envelope of 
\begin{equation}
    \Omega_G(t) = \frac{\pi \left( \exp\left(-\frac{(t - \frac{T}{2})^2}{2\sigma^2}\right) - \exp\left(-\frac{T^2}{8\sigma^2}\right) \right)}{\sqrt{2\pi\sigma^2} \mathrm{erf}\left(\frac{T}{\sqrt{8}\sigma}\right) - T \exp\left(-\frac{T^2}{8\sigma^2}\right)}, 
   \label{eq: DRAG_envelope}
\end{equation}
where $\sigma = T/4$. In frame (a) we explore the robustness with respect to errors in the Rabi frequency. For all solutions the value of $\Omega(t)$ is varied smoothly and the subsequent error is calculated. 
In the case of CP, all $\Omega_n$ are assumed to have the same systematic error. Similarly, in the lower frame we explore the robustness of the different approaches with respect to errors in the qubit frequency. 

We observe that both CP solutions outperform their counterparts by a considerable margin, particularly for small deviations from the ideal conditions. In both frames, their behavior diverges significantly when larger deviations arise. For the robust solution, the optimization included five points near and $\epsilon_{\text{max}} = 0.07$, whereas for the precise solution \(\epsilon_{\max} = 0\); this difference is reflected clearly in the final results. The robust solution exhibits an almost symmetric profile with two peaks at nonzero \(\epsilon\), each surpassing or equaling the DRAG solution over a wide range of deviations. By contrast, the precise solution features a sharply defined peak whose accuracy exceeds that of any other method; however, its lack of robustness likely limits its applicability in most real-world systems. 

\begin{figure}
	\centering 
	\includegraphics[width=0.8\columnwidth, angle=0]{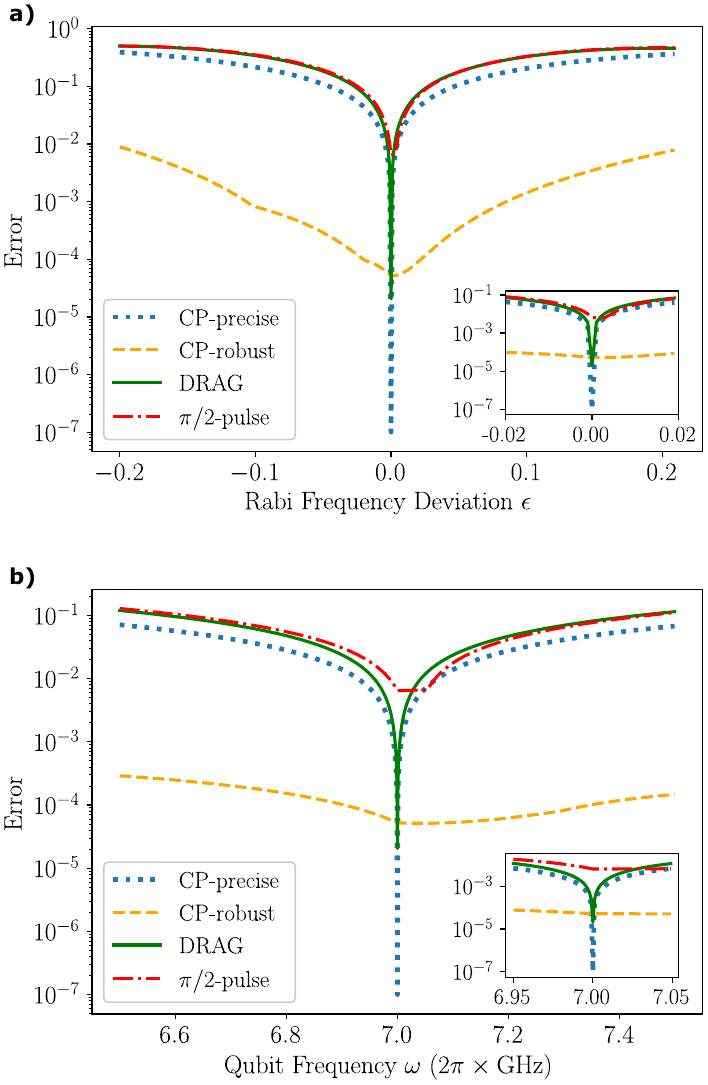}	
	\caption{(a) Average population error in states $\ket{0}$ and $\ket{1}$ as a function of Rabi frequency error $\epsilon$. Results are shown for a single $\frac{\pi}{2}$-pulse,  precise composite sequence, a robust CP, and a DRAG-shaped pulse optimized to maximize transition probability. The inset focuses on the region around $\epsilon = \pm 0.02$.
(b) Average population error in states $\ket{0}$ and $\ket{1}$ as a function of qubit frequency $\omega$. All four pulse types from (a) are included, assuming $\epsilon = 0$ for all. Optimization parameters are $N = 8$, with $\epsilon_{\text{max}} = 0$ for the precise solution and $\epsilon_{\text{max}} = 0.07$ for the robust one.
Once again CP outperforms both $\pi/2$ pulses and DRAG in terms of precision. The robust solution maintains a higher robustness than DRAG, though with slightly lower precision. Notably, the required pulse area (see Table~\ref{table: params}) is lower than the one for a complete population transfer. Overall the main difference compared to Fig.~\ref{fig: pop_trans} is a reduction in accuracy for both CP solutions.} 
	\label{fig: inf}
\end{figure}

The next set of simulations focuses on half population transfer, illustrated in Fig.~\ref{fig: inf}. We again observe performance surpassing both a simple \(\pi/2\) pulse and a DRAG shaped pulse, in terms of both accuracy and robustness. Although the distinction between the two CP solutions remains evident and is even more pronounced in the second plot, it comes at a slight cost in accuracy for the precise solution. Overall, the method performs comparably well on this task, offering heightened robustness at the expense of some accuracy. The underlying reason for this trade-off is not yet clear; we hypothesize that the more complex nature of the operation demands a time-varying \(\Omega(t)\), rendering constant pulses suboptimal. We plan to investigate this further in a future work. Nonetheless, the method is still shown to produce viable solutions that can be applied to real world systems independent of coherent errors and system parameters.

Additional materials exploring the robustness of all simulations to systematic errors in both the anharmonicity and qubit frequency can be found in Appendix~\ref{app: B}.



\subsection{Quantum gates}
\begin{figure}
	\centering 
	\includegraphics[width=0.8\columnwidth, angle=0]{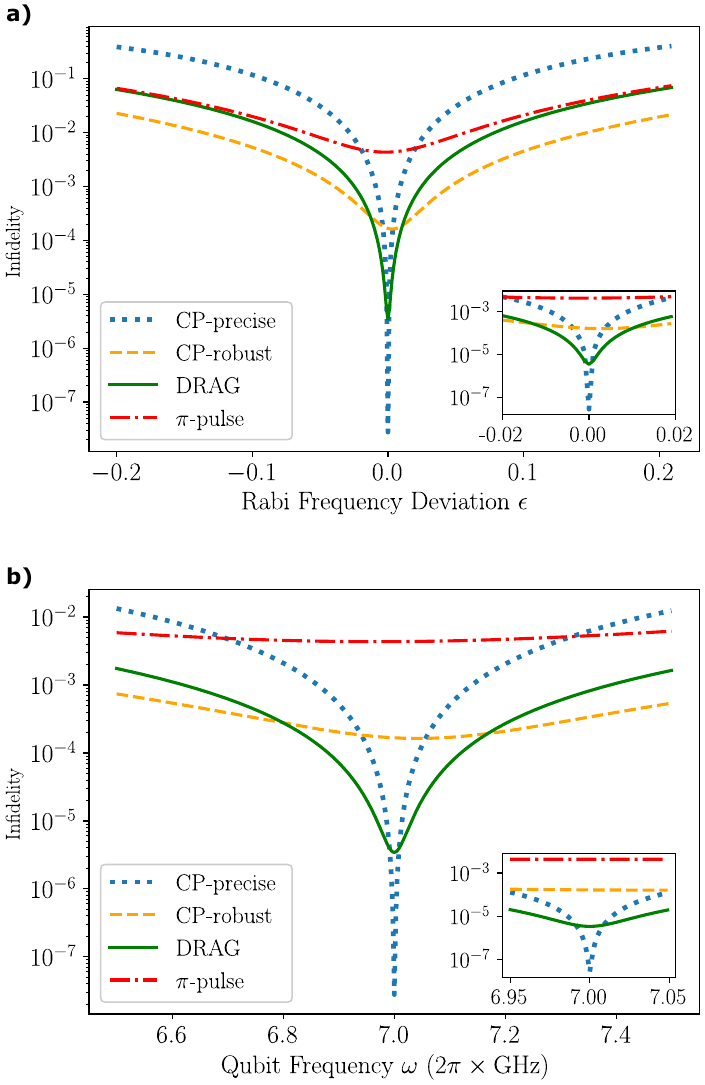}	
	\caption{(a) Infidelity of the $X$ gate as a function of
Rabi frequency error $\epsilon$. Precise CP
sequence, a robust CP, a DRAG shaped pulse optimized to reduce infidelity, and a single $\pi$-pulse. The inset zooms in on the region around $\epsilon = \pm 0.02$.
(b) Infidelity of the $X$ gate as a function of qubit frequency $\omega$, for the same four solutions.
The optimization parameters are $N = 8$, $\epsilon_{\text{max}} = 0$ for CP-precise and $\epsilon_{\text{max}} = 0.07$
CP-robust. Apart from the two minima observed in the robust case, all methods exhibit trends similar to those in Fig.~\ref{fig: pop_trans}, albeit with reduced overall precision. This is mostly true for CP-robust.}
\label{fig: gate_fidelities}
\end{figure}

\begin{figure}
	\centering 	\includegraphics[width=0.8\columnwidth, angle=0]{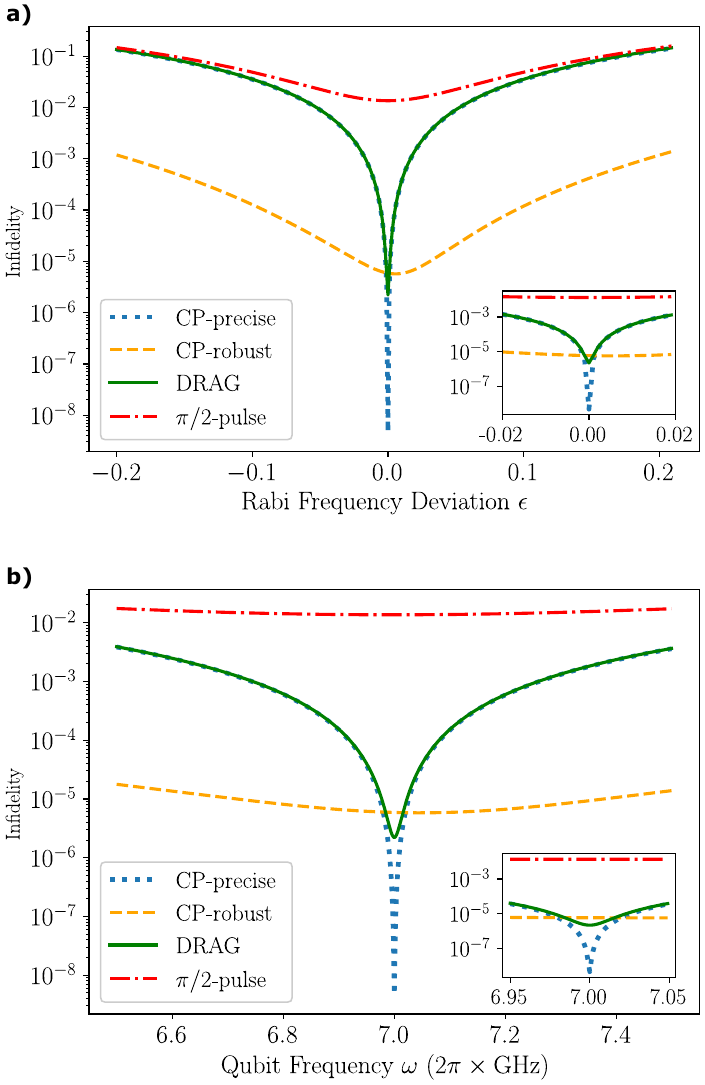}	
	\caption{(a) Infidelity of the $\sqrt{X}$ gate as a function of
Rabi frequency error $\epsilon$. Precise CP
sequence, a robust CP, a DRAG shaped pulse optimized to reduce infidelity, and a single $\pi$-pulse. The inset zooms in on the region around $\epsilon = \pm 0.02$.
(b) Infidelity of the $\sqrt{X}$ gate as a function of qubit frequency $\omega$, for the same four solutions.
The optimization parameters are $N = 8$, $\epsilon_{\text{max}} = 0$ for CP-precise and $\epsilon_{\text{max}} = 0.07$
CP-robust.} 
	\label{fig: sqrtx}
\end{figure}

We follow the same approach to produce high fidelity error-protected quantum gates. We apply our method to the $X$ and $\sqrt{X}$ gates. The optimal phases
and Rabi frequencies are found by maximizing the fidelity function \cite{Pedersen_2007}
\begin{equation}
   F(\{\Omega_k\},\{\phi_k\}) = \frac{\mathbf{Tr}(M_{rel}M^{\dagger}_{rel}) + |\mathbf{Tr}(M_{rel})|^2}{n_{rel}(n_{rel}+1)},
   \label{eq: Fidelity}
\end{equation}
where $M_{rel} = P U^{\dagger}_0 U P$ and $P$ is the projection operator on to the relevant qubit subspace with dimension $n_{rel}=2$, $U_0$ is the target gate and $U$ is the actual unitary.

Once again we favour pulses with smaller areas, and compare our method with a $\pi$ and DRAG pulse, this time optimized to minimize \eqref{eq: Fidelity}. Results for precise and robust solutions of the $X$ gate are shown in Fig.~\ref{fig: gate_fidelities} and the parameters of the gates are given in the lower half of Table ~\ref{table: params}. We see that in the case of operators the CP-method performs slightly worse than in the simpler case of population transfer, which is expected as the task now also takes into account the phase of the final propagator. More on which factors limit the accuracy of the gate can be seen in Table \ref{table: split_gates}. Nevertheless, the significant difference between CP and DRAG remains apparent, though this time less in terms of robustness. In addition, for large errors, the robust solution still outperforms the other approaches. 

The results for the $\sqrt{X}$ gate (Fig.~\ref{fig: sqrtx}) closely resemble those of the $X$ gate, albeit with improved precision across both CP solutions. Notably, the CP-robust sequence achieves approximately a twofold enhancement in precision. Overall, CP methods demonstrate a distinct advantage over the alternative approaches for this problem.

Table~\ref{table: split_gates} provides a detailed breakdown of the infidelity contributions for both $\sqrt{X}$ and $X$ gates implemented via CP sequences, categorized into three error types. The dominant source of error is \textit{leakage}, given by $\mathbf{Tr}(M_{rel}M_{rel}^{\dagger})$, followed closely by \textit{rotational error}, which we get by computing $|\mathbf{Tr}(M_{rel})|$. Their nearly equal magnitudes, for all gates, should be expected, as Eq.~\eqref{eq: Fidelity}, corresponds to a weighted average of these two components, meaning that a significant difference between the two would lead to and increased infidelity. The third error metric is \textit{phase error}, defined in \cite{shill}. It is inherently included within the rotational error calculation; however, by computing it separately, we can verify that, in most cases, its impact remains minor relative to the other sources of error.

Lastly, it is worth noting that the parameters for CP-precise in both Fig.~\ref{fig: sqrtx} and Fig.~\ref{fig: inf} are identical (see Table~\ref{table: params}), illustrating that quantum gate sequences can also be utilized for population transfers. However, this is not a general principle. In the case of complete population transfer (Fig.~\ref{fig: pop_trans}), the sequence is specifically optimized to transfer population from $\ket{0}$ to $\ket{1}$ and performs poorly as an implementation of an $X$ gate. Additional materials, including an analysis of the state vector trajectories under various CP sequences, which further support this observation, are provided in~\cite{megit}.


\section{Summary and conclusions}

In this work we have described a method for quantum
control in transmon qubits, which suppresses errors from
various sources. Specifically, we show that our approach
is capable of reducing the errors, due to (i) population leakage outside of the computational basis, (ii) deviations in the control parameters, and (iii) variations in both qubit frequency and anharmonicity of the system. This
is achieved by using a sequence of pulses with suitably
chosen Rabi frequencies and relative phases. As demonstrated in this work, the
composite pulse scheme can potentially cancel transitions to higher
states, make the errors more robust and at the same time shorten
the pulse duration. The former
leads to a higher accuracy, while the latter allows for
larger circuit depth, which is limited by the coherence
time of the system. Moreover, employing robust composite sequences significantly enhances resilience against so-called coherent errors, including deviations in the Rabi frequency, qubit frequency, and anharmonicity.

Finally, our technique can be seen as an alternative
to the popular method of DRAG pulses \cite{DRAG}, which is
currently the standard approach to leakage reduction. 
While both methods provide high fidelity, our approach also improves robustness against coherent errors, making it a more flexible solution in the presence of imperfections in control parameters. We also hypothesize that our approach can be applied on to different non-linear superconducting qubits such as flux or fluxonium qubits \cite{Rastelli_2015}.

\section*{Acknowledgements}
 This research is supported by the Bulgarian national plan for recovery and resilience, Contract No. BG-RRP-2.004-0008-C01 (SUMMIT), Project No. 3.1.4 and by the European Union’s Horizon Europe research and innovation program under Grant Agreement No. 101046968 (BRISQ).

\appendix
\section{Deriviation of the Driven Transmon Hamiltonian} \label{app: A}
From Eq.~\eqref{eq: n_sum}, we know that the Hamiltonian in Eq.~\eqref{eq:Hamiltonian tr} in the charge basis can be expressed as
\begin{equation}
(\hat{H}_{\mathrm{tr}})_{j,k} = 4E_C (-n_{\text{cut}} + j)^2 \delta_{j,k} - \frac{1}{2} E_J (\delta_{j+1,k} + \delta_{j-1,k}),
    \label{eq: matrix_ham rep}
\end{equation}
where \( n_{\text{cut}} = 30 \) is the parameter that controls the dimensionality of the system and the matrix indices range from \( j_{\min}, k_{\min} = 0 \) to \( j_{\max}, k_{\max} = 2n_{\text{cut}} \).

Now, given Eq.~\eqref{eq: matrix_ham rep}, we diagonalize the Hamiltonian and obtain the transition matrix \( P \), which transforms the charge basis into the energy basis. The columns of \( P \) consist of the eigenvectors of \( \hat{H}_{\mathrm{tr}} \). Using this transformation, we can express the charge operator in the energy basis, where it takes the form
\begin{equation}
    \hat{n}' = P^{\dagger} \hat{n} P =   \sum_{l=0}^{2n_{\text{cut}}} \sum_{j=0}^{2n_{\text{cut}}} \lambda_{j,l} \ket{j} \bra{l} + \text{h.c.},
    \label{eq: transformed n}
\end{equation}
with \(\ket{i}\) denoting the \(i\)-th energy eigenstate and ``h.c.'' standing for Hermitian conjugate. The dimensionless parameters $\lambda_{i,j}$, which are derived numerically and satisfy  
$
\lambda_{i,j}^{*} = \lambda_{j,i} \quad \text{and} \quad \lambda_{j,j} = 0,
$  
define the strength of the transition $\ket{i} \to \ket{j}$.

Using Eq.~\eqref{eq: transformed n}, we incorporate Eq.~\eqref{eq:controled ham} into \( \hat{H}_{\mathrm{tr}} \) to obtain the full driven Hamiltonian in the energy basis
\begin{equation}
    \hat{H}(t) = \left( \sum_{j=0}^{2n_{\text{cut}}} e_j \Pi_j \right) + \Omega(t) \cos(\omega_d t + \phi) \hat{n}',
    \label{eq: full_ham_energy}
\end{equation}
where \( \Pi_j = \ket{j} \bra{j} \) is the projector onto the \( j \)-th energy level and $e_j$ is the $j$-th energy of the system. 

We are now ready to move to a rotating frame at the driving $\omega_d$. The transformation is defined by the matrix
\begin{equation}
    R(t) = \sum_{j=0}^{2n_{\text{cut}}} \exp(-ij\omega_d t)\, \Pi_j,
    \label{eq:rotating}
\end{equation}
which determines the transformed Hamiltonian according to
\begin{equation}
    \hat{H}^{R}(t) = R^{\dagger}(t) H R(t) + i\dot{R}^{\dagger}(t) R(t).
    \label{eq:transform}
\end{equation}
We now compute this expression for the two parts of Eq.~\eqref{eq: full_ham_energy} separately. Firstly, it is straightforward to see that the second term in Eq.~\eqref{eq:transform} leads to a shift in the diagonal elements, and is given by
\begin{equation}
\dot{R}^\dagger(t)\,R(t)
  =\left[
\sum_{j=0}^{2n_{\mathrm{cut}}} i\,j\,\omega_d
      e^{i\,j\,\omega_d\,t}\,\Pi_j^\dagger
  \right]
  \left[
\sum_{k=0}^{2n_{\mathrm{cut}}} 
      e^{-\,i\,k\,\omega_d\,t}\,\Pi_k
  \right]
\end{equation}
\begin{equation}
=\sum_{j=0}^{2n_{\mathrm{cut}}}\,\sum_{k=0}^{2n_{\mathrm{cut}}}
    ij\,\omega_{d}\;
    e^{i\,(j-k)\,\omega_{d}\,t}\Pi_{j}^\dagger\,\Pi_{k} = \sum_{j=0}^{2n_{\mathrm{cut}}}ij\,\omega_d\;\Pi_j,
    \label{eq:rhr}
\end{equation}
where we have used $\Pi_j^\dagger \Pi_k = \Pi_j \Pi_k = \delta_{j,k} \Pi_j.
$ 

Let us now compute the first term in Eq.~\eqref{eq:transform}. For the diagonal part of Eq.~\eqref{eq: full_ham_energy} we get
\begin{align}
  &\left(\sum_{j=0}^{2n_{\mathrm{cut}}}
         e^{+i j \omega_d t} \Pi_j \right)
     \left(\sum_{m=0}^{2n_{\mathrm{cut}}}
         e_m \Pi_m \right) \left(\sum_{k=0}^{2n_{\mathrm{cut}}}
         e^{-i k \omega_d t} \Pi_k \right) = \notag \\
  &= \sum_{j=0}^{2n_{\mathrm{cut}}} \sum_{m=0}^{2n_{\mathrm{cut}}} \sum_{k=0}^{2n_{\mathrm{cut}}} e^{+i (j-k) \omega_d t} e_m \Pi_j \Pi_m \Pi_k  = \sum_{m=0}^{2n_{\mathrm{cut}}} e_m \Pi_m. 
  \label{eq: diag_tr}
  \end{align}
Clearly, the term remains invariant under the change of basis, which together with Eq.~\eqref{eq:rhr} implies that the transformation modifies the diagonal elements only by subtracting terms proportional to the drive frequency. As a side note, we can set $e_0=0$ since a constant energy term can always be added to the Hamiltonian without affecting the system's dynamics. 

We have now addressed two out of three terms of the transformed Hamiltonian and continue with the last part of Eq.~\eqref{eq: full_ham_energy}
\begin{align}
\hat{n}_{\text{R}}   & = 
\left( \sum_{k=0}^{2n_{\mathrm{cut}}} e^{i k \omega_d t} \Pi_k \right) \notag \\
&\quad \times \left( \sum_{l=0}^{2n_{\text{cut}}} \sum_{j=0}^{2n_{\text{cut}}} \lambda_{j,l} \ket{j} \bra{l} + \text{h.c.} \right) \notag \\
&\quad \times \left( \sum_{m=0}^{2n_{\mathrm{cut}}} e^{-i m \omega_d t} \Pi_m \right),
\label{eq:longa}
\end{align}
where $ \hat{n}_{\text{R}} = R^{\dagger}(t) \hat{n}' R(t)$.
In order to compute this expression we will mainly rely on the following property of the projection operator
\begin{equation}
\Pi_k |j\rangle = \delta_{k,j} |j\rangle,
\end{equation}
which allows us to expand the parenthesis in Eq.~\eqref{eq:longa} and results in
\begin{align}
&\sum_{l=0}^{2n_{\mathrm{cut}}} 
\sum_{j=0}^{2n_{\mathrm{cut}}}
\lambda_{j, l} e^{i j \omega_d t} |j\rangle \langle l| e^{-i l \omega_d t} + \text{h.c.} =  \\
&=\sum_{l=0}^{2n_{\mathrm{cut}}} 
\sum_{j=0}^{2n_{\mathrm{cut}}}
\lambda_{j, l} e^{i (j-l) \omega_d t} |j\rangle \langle l| + \text{h.c.} = \hat{n}_{\text{R}}.
\label{eq:rnr}
\end{align}
Summing up the results by combining Eq.~\eqref{eq:rhr}, Eq.~\eqref{eq: diag_tr} and Eq.~\eqref{eq:rnr}, for the Hamiltonian of a driven transmon in a rotating frame with frequency $\omega_d$ we get
\begin{equation}
\begin{split}
    \hat{H}^{R}(t) & = \sum_{m = 0}^{2n_{\text{cut}}} (e_m - m\omega_d) \Pi_m + \frac{\Omega(t)}{2}[e^{-i(\omega_dt + \phi)} + \text{h.c.}] \\
    &\times \sum_{l=0}^{2n_{\mathrm{cut}}} 
\sum_{j=0}^{2n_{\mathrm{cut}}}
\lambda_{j, l} e^{i (j-l) \omega_d t} |j\rangle \langle l| + \text{h.c.}
\label{eq:rotating_h}
\end{split}
\end{equation}
We are now in a position to apply the rotating wave approximation (RWA). To do so, we assume that the driving frequency $\omega_d$ is significantly larger than any other characteristic time-dependent variation in the system. Additionally, we impose the condition that the magnitude of the driving term satisfies $|\Omega(t)| \ll |\hat{H}^{R}(t)|$. 

These assumptions justify neglecting rapidly oscillating terms, as their contribution averages out over time. More specifically, terms containing exponentials of the form $e^{i(j-l)\omega_d t} e^{\pm i\omega_d t} \neq 1$ integrate to zero over sufficiently long timescales. 

After applying RWA the Hamiltonian then simplifies to
\begin{equation}
\begin{split}
    \hat{H}^{R}(t) & = \sum_{j = 0}^{2n_{\text{cut}}} \mu_j\Pi_j +  
\sum_{j=1}^{2n_{\mathrm{cut}}}
\lambda^{'}_{j} \frac{\Omega_R(t)}{2} |j\rangle \langle j - 1| + \text{h.c.},
\label{eq:rotating_h_rwa}
\end{split}
\end{equation}
where $\mu_j = e_j - j\omega_d$, $\lambda^{'}_j =  \lambda_{j,j-1}$, and $\Omega_R(t) = \Omega(t)e^{-i\phi}$. 

We see that interactions between energy levels that are not nearest neighbors  are removed by the time-averaging process due to their high-frequency oscillations. Moreover, this result mirrors the formulation for a general nonlinear oscillator presented in \cite{DRAG}.

\section{2d plots of infidelity as a function of qubit frequency and anharmonicity} \label{app: B}
\begin{figure}
	\centering 	\includegraphics[width=1.0\columnwidth, angle=0]{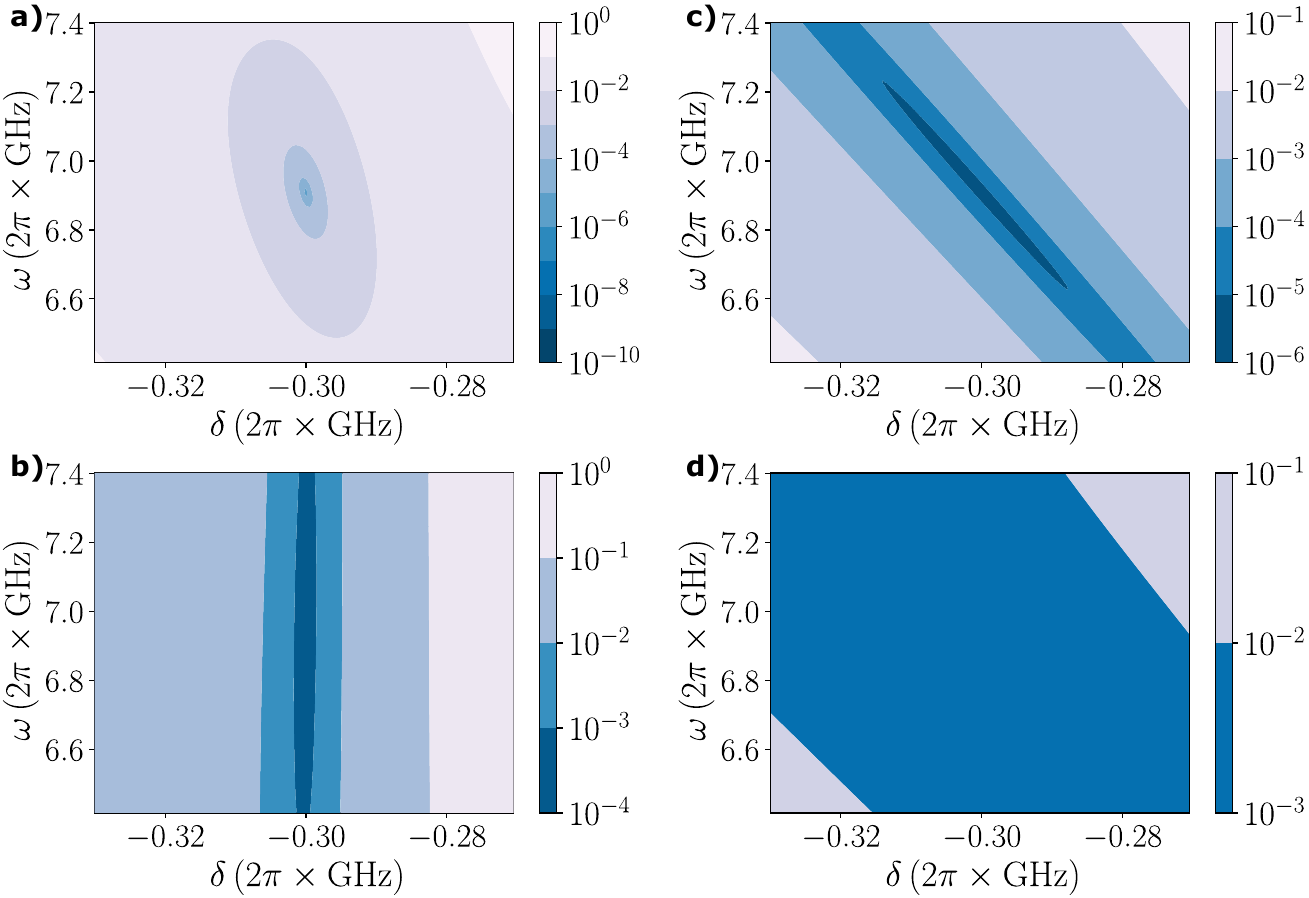}	
	\caption{ Population error in state $\ket{1}$ as a function of anharmonicity $\delta$ and qubit frequency $\omega$ for a complete population transfer: 
(a) Precise CP sequence with $\epsilon_{\max} = 0$, $N = 7$;  
(b) Robust CP sequence with $\epsilon_{\max} = 0.07$, $N = 8$;  
(c) DRAG-shaped pulse;  
(d) Single $\pi$-pulse.  
All plots assume no Rabi frequency error, except for (b), where $\epsilon = -0.047$ is used. The panels exhibit distinct behaviors: (a) shows equal sensitivity to both error types, forming a sharp peak; (d) has a nearly uniform error profile; (b) and (c) show line-shaped profiles—centered at $\Delta \delta = 0$ in (b), and along $\Delta \omega \propto \Delta \delta$ in (c) (where $\Delta \omega \text{ and } \Delta \delta$ are the errors in their respective variables). }
    \label{fig: 2d pop trans}
\end{figure}

\begin{figure}
	\centering 	\includegraphics[width=1.0\columnwidth, angle=0]{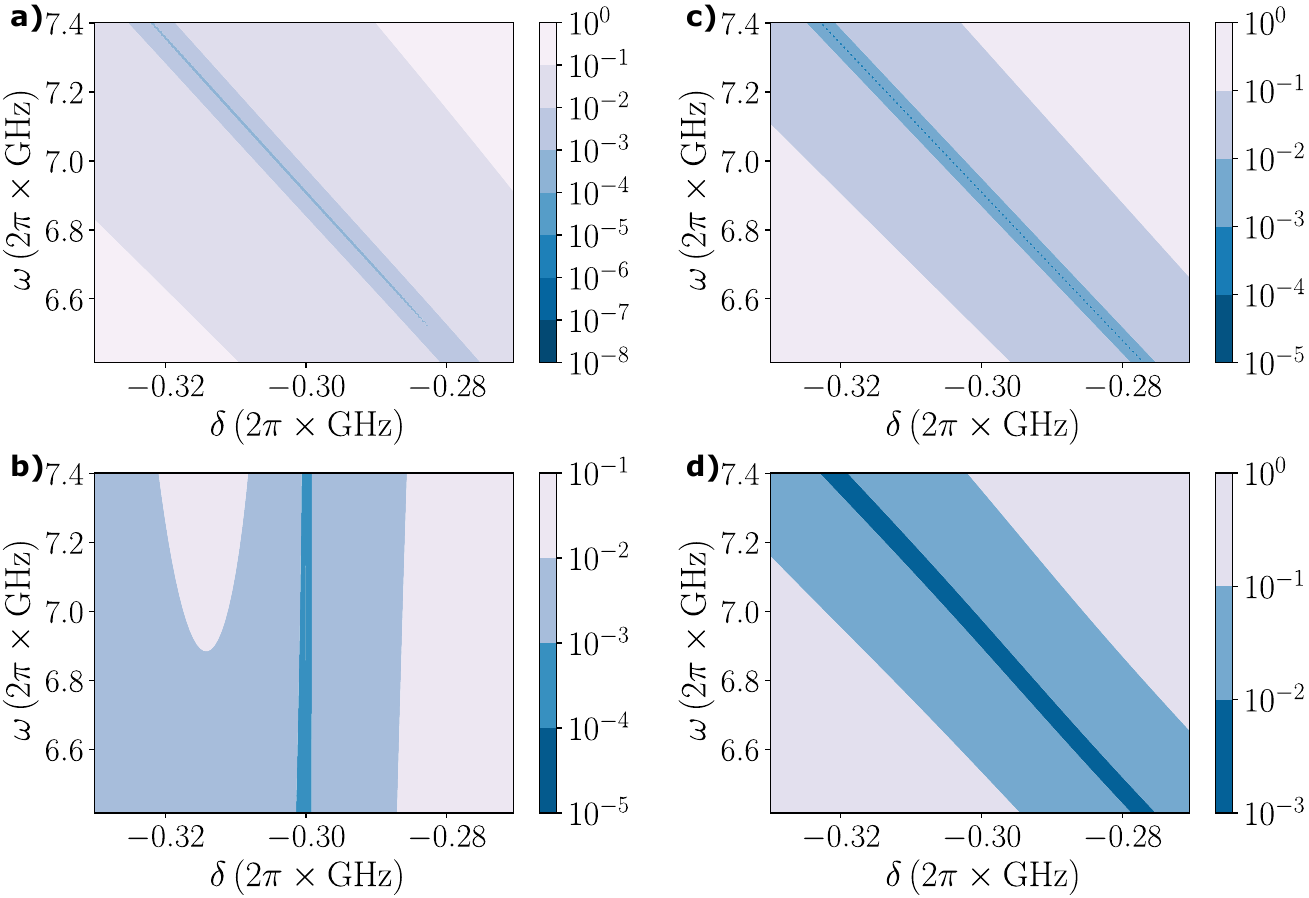}	
		\caption{ Average population error in states $\ket{1}$ and $\ket{0}$, as a function of anharmonicity $\delta$ and qubit frequency $\omega$ for a half population transfer:  
(a) Precise CP sequence with $\epsilon_{\max} = 0$, $N = 8$;  
(b) Robust CP sequence with $\epsilon_{\max} = 0.07$, $N = 8$;  
(c) DRAG-shaped pulse;  
(d) Single $\frac{\pi}{2}$-pulse.  
All plots assume no Rabi frequency error. The overall error profiles in panels (b) and (c) closely resemble those observed in the complete population transfer case. Interestingly, panels (a) and (d) now display behavior similar to that of the DRAG pulse, though with a different level of robustness. In general, the half-population transfer solutions exhibit reduced robustness compared to their full population transfer counterparts.
}
    \label{fig: 2d half pop trans}
\end{figure}

\begin{figure}
	\centering 	\includegraphics[width=1.0\columnwidth, angle=0]{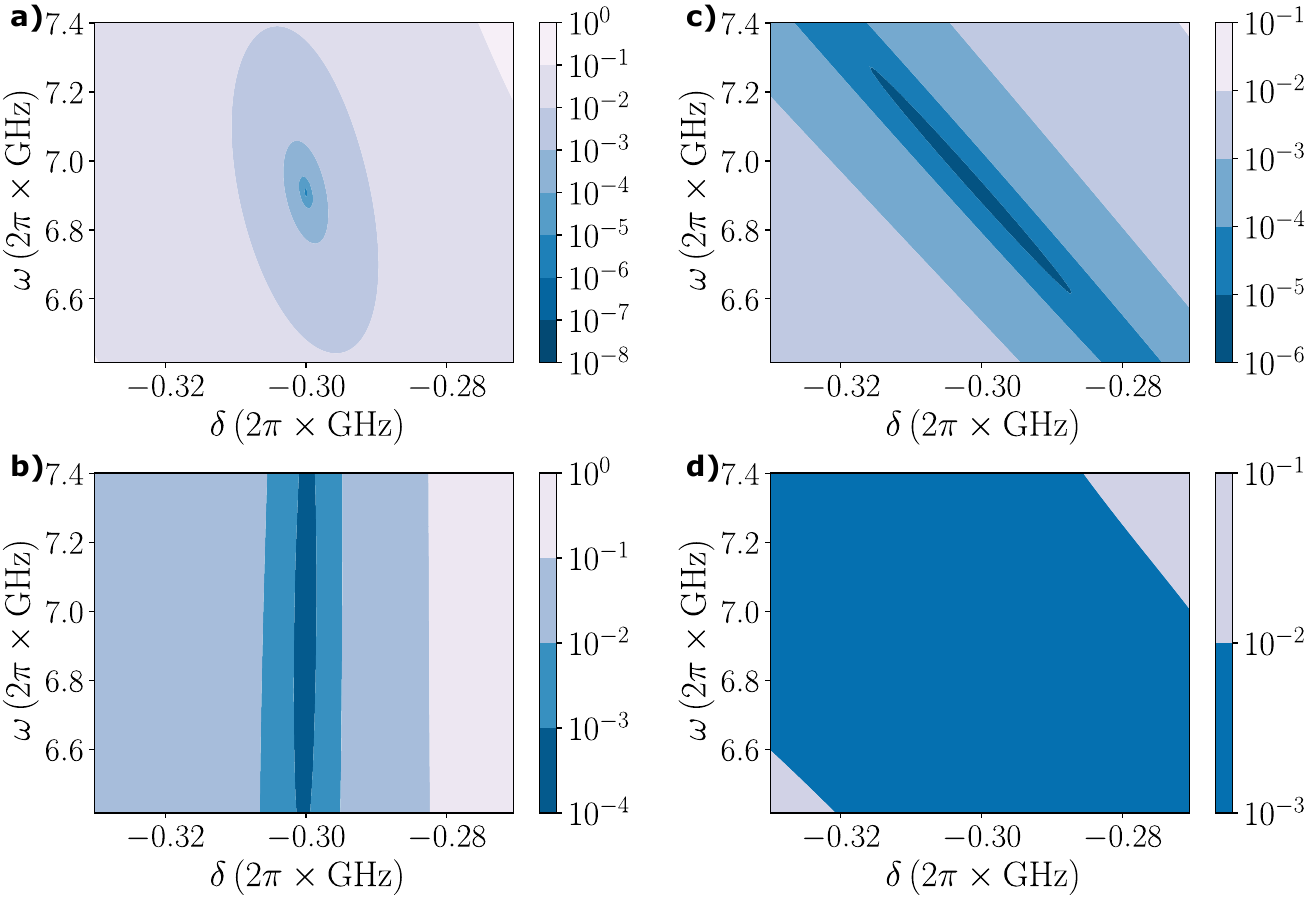}	
		\caption{ 
        Infidelity of the $X$ gate as a function of anharmonicity $\delta$ and qubit frequency $\omega$, for:
    a) Precise CP sequence where 
    $\epsilon_{\max}=0$ and $N = 8$. 
    b) Robust CP sequence where $\epsilon_{\max}=0.07$ and $N = 8$ . 
    c) The DRAG-shaped pulse. 
    d) A single $\pi$-pulse. 
All plots assume no Rabi frequency error.}
    \label{fig: 2d X gate}
\end{figure}

\begin{figure}
	\centering 	\includegraphics[width=1.0\columnwidth, angle=0]{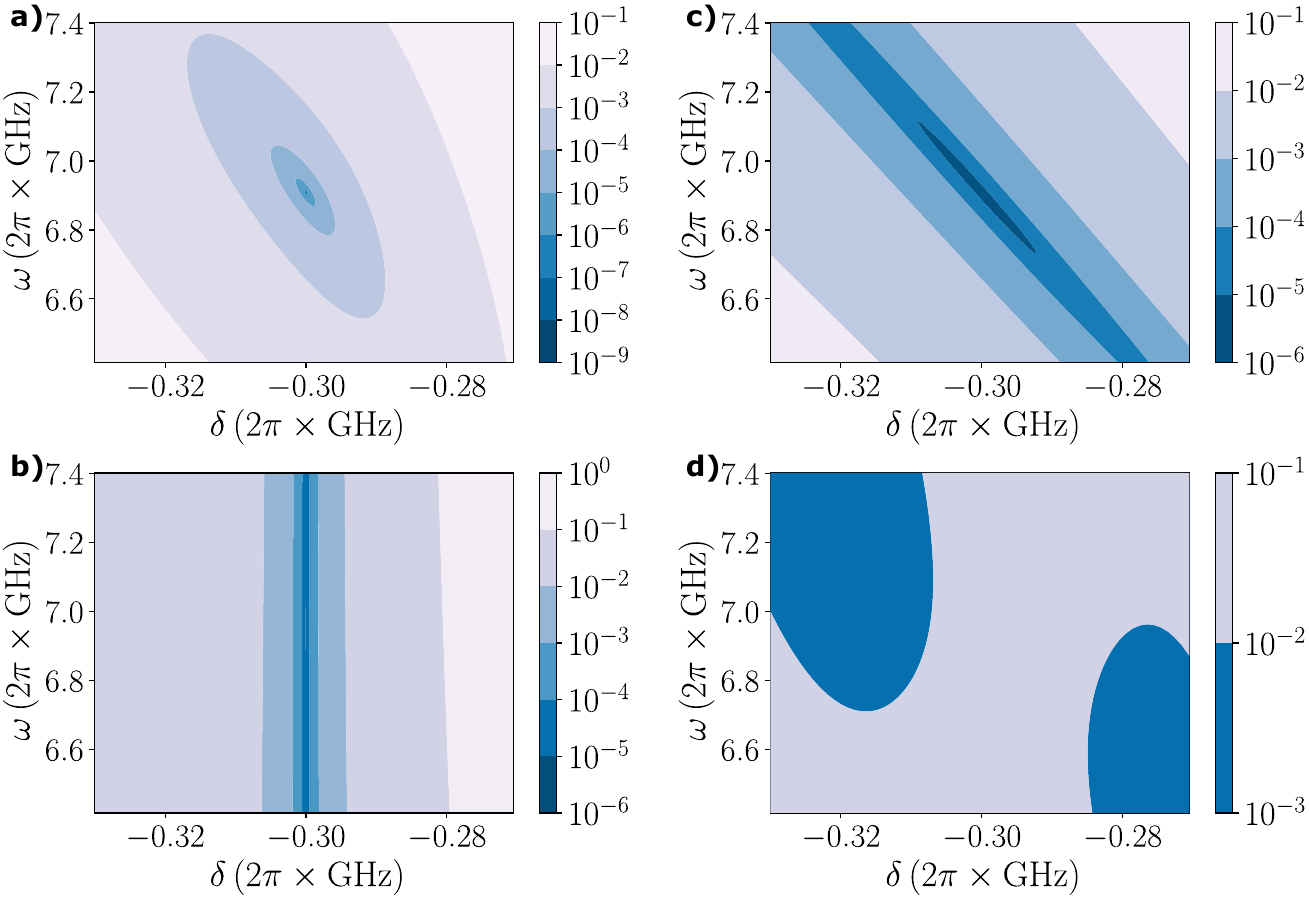}	
		\caption{ Infidelity of the $\sqrt{X}$ gate as a function of anharmonicity $\delta$ and qubit frequency $\omega$, for:
    (a) Precise CP sequence with $\epsilon_{\max} = 0$, $N = 8$;  
(b) Robust CP sequence with $\epsilon_{\max} = 0.07$, $N = 8$;  
(c) DRAG-shaped pulse;  
(d) Single $\frac{\pi}{2}$-pulse.  All plots assume no Rabi frequency error.
    }
    \label{fig: 2d sqrtX gate}
\end{figure}

Figures~\ref{fig: 2d pop trans},~\ref{fig: 2d half pop trans},~\ref{fig: 2d X gate}, and~\ref{fig: 2d sqrtX gate} illustrate the performance of the four approaches in the presence of errors in the system's anharmonicity and qubit frequency. Most CP solutions exhibit broadly similar characteristics across the 4 tasks.

The precise CP solutions, in particular, are similarly sensitive to both types of errors. Their performance profile resembles a sharp peak, where even small deviations from the ideal parameters result in a significant drop in fidelity. The one exception being CP-precise for a half population transfer. There if we define $\Delta \omega$ as the deviation of $\omega$ from its assumed value in the optimization procedure, and similarly $\Delta \delta$ for the anharmonicity, we get a optimal value sensitive to deviations away from the line $\Delta \omega \propto \Delta \delta$. This is not limited strictly to CP-precise, as we can see in Fig.~\ref{fig: 2d half pop trans}, and seems to be a feature of most of the tested approaches.

The robust CP sequences, on the other hand, exhibit strong stability with respect to qubit frequency errors for all cases, and maintain a reasonable performance for anharmonicity errors below 2--3\%. DRAG pulses, by contrast, tend to perform best when the two types of errors are linearly proportional to each other, i.e., when $\Delta \omega \propto \Delta \delta$. Taking this distinction into account could influence which approach an experimentalist might prefer for transmon control, depending on the dominant error behavior in their setup.

Finally, $\pi$ and $\frac{\pi}{2}$ pulses, as expected, exhibit relatively uniform error profiles. This is also true for Fig.~\ref{fig: 2d sqrtX gate}, where the overall error remains close to $10^{-2}$, and thus a slight variation results in a visually distinct plot, despite the absolute value of the error remaining approximately constant. Again the only exception being Fig.~\ref{fig: 2d half pop trans}, even though a pronounced higher robustness compared to the other $3$ approaches is still present.

\FloatBarrier

\end{document}